\documentstyle[12pt,openbib,epsfig,axodraw]{article}

\input epsf
\def\br{\begin{eqnarray}}
\def\er{\end{eqnarray}}
\def\be{\begin{equation}}
\def\ee{\end{equation}}
\def\l{\label}
\def\a{\alpha}
\def\b{\beta}
\def\c{\chi}

\def\L{\Lambda}
\def\S{\Sigma}
\def\n{\eta}
\def\g{\gamma}
\def\G{\Gamma}

\def\m{\mu}
\def\P{\Pi}

\def\d{\delta}
\def\<{\left\langle}
\def\>{\right\rangle}
\def\gc{\<{\frac{\alpha_s}{\pi}}G^{\mu\nu}G_{\mu\nu}\>}
\def\fc{\<{ - \bar{\psi}}\psi\>}
\begin{document}
\vspace{1cm}
\begin{center}
{\large\bf Relating the Quark and Gluon Condensates Through
the QCD Vacuum Energy}\\[.4cm]       
E.~V.~Gorbar\dag ~\footnotemark
\footnotetext{e-mail: gorbar@ift.unesp.br}\hspace{2mm} and
A.~A.~Natale~\footnotemark \footnotetext{e-mail: natale@ift.unesp.br} \\[.2cm]
Instituto de F\'{\i}sica Te\'orica,                                         
Universidade Estadual Paulista\\
Rua Pamplona, 145, 01405, S\~ao Paulo, SP, Brazil
\end{center}
\thispagestyle{empty}
\vspace{1cm}

\begin{abstract}
Using the Cornwall--Jackiw--Tomboulis effective potential
for composite operators we compute the QCD vacuum energy as
a function of the dynamical
quark and gluon propagators, which are related to their
respective condensates as predicted by the operator product
expansion. The identification of this result to the vacuum
energy obtained from the trace of the energy-momentum tensor
allow us to study the gluon self-energy, verifying that it
is fairly represented in the ultraviolet by the
asymptotic behavior predicted by the operator product expansion, and in the
infrared it is frozen at its asymptotic value at one scale of the order
of the dynamical gluon mass. We also discuss the implications
of this identity for heavy and light quarks. For heavy
quarks we recover, through the vacuum energy calculation, the relation
$m_{f} <{\bar{\psi_f}}\psi_f> \sim - \frac{1}{12} \gc$ obtained
many years ago with QCD sum rules.

\end{abstract}

\vspace{5mm}

\dag On leave of absence from Bogolyubov Institute for Theoretical
Physics, Metrologichna 14-b, 252143, Kiev, Ukraine 
\newpage

\section{Introduction} 

\noindent
The QCD vacuum energy can be computed
through the use of the effective potential for composite
operators~\cite{cornja}. This vacuum energy is a
function of the masses of the theory~\cite{cornja,cornor}.
For quarks the effective mass is a sum
of the running mass and the dynamical one and the quark
condensate is an order parameter for chiral symmetry
breaking (see, {\it e.g.},~\cite{kiev}). According to
the operator product expansion (OPE)~\cite{pol,pasc}
the dynamical mass is a function of the quark condensate. 
Therefore, the vacuum energy 
involves the bare quark masses and the quark condensate.

The concept of a dynamical gluon mass~\cite{corn} is
crucial to calculate the QCD vacuum energy. (We would
like to emphasize that the presence of a dynamically
generated mass does not mean that gluons can be
considered as massive asymptotic states like
dynamically generated quark mass does not mean that quarks can be
observed as massive asymptotic states. Why quarks and gluons
are not observed as free states is the well known problem 
of confinement). As discussed
at length in Ref.~\cite{corn}, the vacuum energy is
finite as long as the gluon mass decreases at large
momentum, and it vanishes when the gluon mass is zero.
The infrared divergences that plagued the effective
potential calculations are also absent for a finite
dynamical gluon mass. Finally, the gluon mass is related to
the gluon condensate, and a precise relation, as in
the case of the dynamical quark mass, has been obtained
through the OPE~\cite{lav}. It is also known that the
gluon contribution dominates the vacuum energy. However, not 
much is known about the infrared behavior of the dynamical gluon
mass. 

On the other hand the vacuum energy can also be obtained
through the trace of the energy momentum tensor and is
a function of the quark and gluon condensates and the
bare quark mass. Equalizing these two expressions
we obtain information about the dynamical masses of the
theory and the condensates.
In this work we use two different calculations
of the vacuum energy density in order to determine the behavior of the
dynamical masses, particularly of the gluon dynamical mass. 
We show that it
is fairly represented in the ultraviolet by the
asymptotic behavior predicted by the OPE, and in the infrared it
is equal to its value at a scale of the order
of the dynamical gluon mass. 
Within this approach we are also able to
obtain a relation between the quark and gluon condensates in the
heavy quark limit. It is oportune to recall that many 
years ago Shifman {\it et
al.} introduced the notion of a nonvanishing gluon condensate~\cite{shif1}.
Making use of the OPE they
started the successful program of QCD sum rules, where observable
quantities can be written as a function of quark and
gluon condensates, and one of their main
results is the relation between quark an gluon condensates
valid in the heavy quark limit~\cite{shif2}:
\be
m_{f} <{\bar{\psi_f}}\psi_f> \sim - \frac{1}{12} \gc .
\l{qgcon}
\ee
This relation emerges naturally in our calculation of
the vacuum energy. We also discuss what can be learned
from our approach in the chiral and light fermion limits.
In Section 2 we setup
the equations for the vacuum energy. Section 3 contains
a discussion of our computational ansatze for the dynamical
masses, and Section 4 contains the vacuum energy calculation. In the
last section we discuss the method and present 
the conclusions.

\section{The effective potential for composite operators and
the QCD vacuum energy}       

\noindent

For a non-Abelian gauge theory the effective potential
has the form~\cite{cornja}              
\br
V(S,D,G) &=& - \imath \int \frac{d^4p}{(2\pi)^4} 
Tr ( \ln S_0^{-1}S - S_0^{-1}S + 1) \nonumber \\
&& - \imath \int \frac{d^4p}{(2\pi)^4}
Tr ( \ln G_0^{-1}G - G_0^{-1}G + 1) \nonumber \\
&& + \frac{\imath}{2} \int \frac{d^4p}{(2\pi)^4}
Tr ( \ln D_0^{-1}D - D_0^{-1}D + 1) \nonumber \\
&& + V_2(S,D,G), 	
\l{vfull}                                                              
\er                                                        
where $S$, $D$ and $G$ are the complete propagators of
fermions, gauge bosons, and Faddeev-Popov
ghosts, respectively; $S_0$, $D_0$, and $G_0$ the corresponding bare
propagators. $V_2(S,D,G)$ is the sum of all two-particle
irreducible vacuum diagrams,  depicted in Fig.1, and 
the equations   
\be
\frac{\d V}{\d S}=\frac{\d V}{\d D}=\frac{\d V}{\d G}=0,
\l{delv}
\ee
are the Schwinger-Dyson equations (SDE) for fermions, gauge
bosons, and ghosts.
                         
We can represent $V_2(S,D,G)$ analytically by
\br
\imath V_2(S,D,G) &=& - \frac{1}{2} Tr(\G S \G S D)
- \frac{1}{2} Tr (F G F G D) \nonumber \\
&& + \frac{1}{6} Tr(\G^{(3)}D\G^{(3)}DD)
+ \frac{1}{8} Tr(\G^{(4)}DD)
\l{v2full}
\er
where $\G$, $F$, $\G^{(3)}$ and $\G^{(4)}$ are respectively
the proper vertex of fermions, ghosts, trilinear and quartic
gauge boson couplings~\cite{na}. In Eq.(\ref{v2full}) we 
have not written the gauge and Lorentz indices, as well as the
momentum integrals.        

The complete gauge boson propagator $D$ is related
to the free propagator by
\be
D^{-1} = D_0^{-1} - \Pi^T,
\l{dfull}
\ee
where $\Pi^T$ is the gluon polarization tensor,
which is obtained from Eq.(\ref{vfull}) and Eq.(\ref{delv}), 
and described by
\be
\Pi^T = \G S \G S + F G F G 
- \frac{1}{2} \G^{(3)} D \G^{(3)} D - \frac{1}{2} \G^{(4)} D.
\l{picomp}
\ee
The diagrams contributing to $\Pi^T$
are shown in Fig.2, and this
self-energy is the one that admits a massive
solution as described by Cornwall~\cite{corn}. 
The Landau gauge expression for the complete
gluon propagator is
\be
D^{\m \nu}(p^2) = - \frac{\imath}{p^2-\Pi(p^2)} 
\left( g^{\m \nu} - \frac{p^\m p^\nu}{p^2} \right) .
\l{dlprop}
\ee

The complete fermion propagator $S$ is related to the
free propagator by
\be
S^{-1} = S_0^{-1} - \S ,
\l{sdf} 
\ee
where $S_0 = \imath / ( \not \! p - m_0 )$ ($m_0$ is the bare
quark mass) and 
$\S$ is the Schwinger-Dyson
equation for the quark self-energy, which is given by
\be
\S = - \g S \G D,
\l{se}
\ee
where $\g$ is the bare vertex.

The vacuum energy density is given by the effective
potential calculated at minimum subtracted by its
perturbative part, which does not contribute to dynamical
mass generation ~\cite{cornja,cornor,corn}
\be
\< \Omega \> = V_{min}(S,D,G) - V_{min}(S_p,D_p,G_p),
\l{omega}
\ee
where $S_p$ is the perturbative
counterpart of $S$, etc... 

$V_{min}$ is obtained substituting the solutions of
Eq.(\ref{delv}) into Eq.(\ref{vfull}). It is oportune to
recall that we shall compute the vacuum energy by using
ansatze for the dynamical masses. Although these ansatze are consistent
with the OPE they are simple approximations to the full solution
of the Schwinger-Dyson equations. However, since we calculate
the potential at its stationary point,
the result depends significantly less on the
approximations used, as was noticed for the vacuum energy calculation
in the case of fermions~\cite{castorina}. If we use the corresponding
Schwinger--Dyson equations, $V_{min}$ can be
divided in two parts (as will become clear in the following)
\be
V_{min} = V_{min}^{f} + V_{min}^{g},
\l{vmin}
\ee
where the labels $f$ and $g$ indicate the fermionic and
gluonic parts of the vacuum energy. The quark contribution
to the effective potential obtained by substituting the solution
of $\d V / \d S$ into Eq.(\ref{vfull}) and (in Euclidean
space with $P^2 \equiv -p^2$) is equal to~\cite{castorina}
\be
V_{min}^f = 2 N\int \frac{d^4P}{(2\pi)^4} \,
\left[ - \ln ( \frac{P^2 + \S^2}{P^2 + m_0^2} ) +
\frac{\S^2 - \S m_0}{P^2 + \S^2} \right],
\l{vminf}
\ee
where N is the number of colors (quarks are in the
fundamental representation of SU(N)).

To compute $V_{min}^{g}$ we make an approximation
which considerably simplifies the calculation. This approximation
frequently performed to solve the gluon polarization 
tensor is due to Mandelstam~\cite{mand} and
consists in neglecting the ghosts and the
diagram with quartic coupling.  The neglect of ghosts
diagrams was shown to be reasonable, because their
contribution is numerically small~\cite{mand}.
These approximations, with the use of the
Landau gauge, were shown to be satisfactory in 
the lengthy and detailed work of
Brown and Pennington~\cite{brown}. Although it may
seems rough
this approximation has been verified to be phenomenologically
consistent in Ref.~\cite{mont} as long as the vacuum 
polarization tensor involves a mass scale. Furthermore, as can
be easily verified, by expanding the integrand 
in Eq.(\ref{omeg}) in powers of
$\Pi$ we obtain a result similar to the Cornwall determination of the
vacuum energy in the case of a pure gluon theory~\cite{corn}.
Therefore, only the diagram with trilinear
coupling should be considered in Eq.(\ref{v2full}),
{\it i.e.} the gluon polarization tensor is given by
\be
\P^T = - \frac{1}{2} \G^{(3)} D \G^{(3)} D.
\l{piapro}
\ee
With this approximation the two-loop gluon contribution
in Eq(\ref{v2full}) is reduced to
\be 
V^g_2 = \frac{-\imath}{6} Tr(\G^{(3)}D\G^{(3)}DD).
\l{ommand}
\ee
Finally, we obtain ~\cite{mont}
\be
\< \Omega_g \> = - \frac{3(N^2 -1)}{2} \int \frac{d^4P}{(2\pi)^4} \,
\left[ \frac{\P}{P^2+\P} - \ln \left( 1+\frac{\P}{P^2}
\right) + \frac{2}{3} \frac{\P^2}{P^2(P^2+\P)} \right] ,
\l{omeg}
\ee
where all the quantities are in Euclidean space and $N=3$ for
QCD.

Thus, according to Eq.(10), the vacuum energy is
\be
\< \Omega \> = \< \Omega_f \> + \< \Omega_g \> ,
\l{ome}
\ee
where $\<\Omega_f\>$ is (see Eqs. (10) and (12))
\br
\< \Omega_f \> &=& 2N \int \frac{d^4P}{(2\pi)^4} \,
\left[ - \ln \left( \frac{P^2 + \S^2}{P^2 + m_0^2} \right) +
\frac{\S^2 - \S m_0}{P^2 + \S^2} \right] \nonumber \\
&&- 2N \int \frac{d^4P}{(2\pi)^4} \,
\left[ - \ln \left( \frac{P^2 + m_f^2}{P^2 + m_0^2} \right) +
\frac{m_f^2 - m_f m_0}{P^2 + m_f^2} \right] ,
\l{omef}
\er
where $m_f^2 \equiv m_f^2 (P^2)$ is the running quark mass, which is the
perturbative solution of the fermionic Schwinger-Dyson equation
with bare quark mass $m_0$. Note that Eq.(\ref{omef}) has been written
for a unique fermion of mass $m_f$. If we consider $n_f$ fermions of
the same mass it should be obviously multiplied by $n_f$. There is
only one divergence in $\< \Omega_f \>$  that is associated to the bare quark
mass renormalization as  will be discussed afterwards. For gluons with
dynamically  generated mass the gluon contribution to the vacuum energy 
is free of divergences ~\cite{corn,mont}. As long as we
have appropriate expressions for $\S$ and $\Pi$ the calculation
of $\< \Omega \>$ is an easy task.

\section{The self-energy of quarks and gluons } 

\noindent

To compute the vacuum energy we need reasonable ansatze
for the quark and gluon self-energies.
In the case of quarks the high energy 
behavior of the dynamical mass given by the OPE is~\cite{pol,pasc}
\be
m_{dyn} (P^2) \simeq  \frac{3(N^2-1)}{8N^2} g^2(P^2)
\frac{\fc}{P^2} (\ln{P^2/\L_Q^2})^{-\g_m},
\l{sope}
\ee
where $g^2(P^2) $ is the running coupling constant 
\be
g^2(P^2) = \frac{2b}{\ln{P^2/\L_Q^2}},
\l{rung}
\ee
where $b = 24\pi^2/(11N - 2 n_f)$. In the framework of SDE
the same behavior was found in Ref.~\cite{higa}. 
As well known
the asymptotic behavior
of the running quark mass is the following:
\be
m_f^{asymp} ( P^2) \sim  m_f (\m^2) \left( \frac{\ln{P^2/\L_Q^2}}
{\ln{\m^2/\L_Q^2}} \right)^{-\g_m},
\l{run}
\ee
where $\L_Q$ is the  QCD scale, $\g_m = 9 C_2/(11 N - 2n_f)$,
$C_2 = (N^2 - 1)/2N $ is the quadratic Casimir of the fundamental
representation, and $m_f (\m^2)$ is the running quark mass at the scale $\m^2$
(a mass scale where perturbative QCD can be safely applied),
and is related to the bare mass by
\be
m_f(\m) = m_0 (\L) Z_m^{-1} (\m,\L),
\l{mz}
\ee
where $\L$ is a ultraviolet cutoff and $Z_m^{-1} (\m , \L)$ is
a renormalization constant for the bare mass term
\be
Z^{-1}_m (\m,\L) = \left( \frac{\ln{\m^2/\L_Q^2}}
{\ln{\L^2/\L_Q^2}} \right)^{\g_m} .
\l{rem}
\ee
A discussion of the
renormalization procedure applied to the effective
potential for composite operators can be found in Ref.~\cite{bar}.

Eqs.(\ref{sope}) and (\ref{run}) involve the high momentum part of the
self-energy. To establish our ansatz for $\S$ for all $P^2$, we recall 
a few points on the infrared behavior of the quark mass. 
There are several arguments in favor of the relation~\cite{mck}
\be
\S ( P^2 \rightarrow 0) \approx \fc^{1/3}
\l{mdyn0}
\ee
for massless or very light
quarks. For heavy quarks we have
\be
\S( P^2 \rightarrow 0) \approx  m_f(P^2)|_{P^2 = m_f^2}.
\l{mh}
\ee
These equations provide an idea about the ansatz we
should use for the quark self-energy. Actually, there is a full
numerical solution of the quark Schwinger-Dyson equations
with bare
massive quarks performed by Maris and Roberts~\cite{maro}, and
they verify that for very heavy quarks the self-energy is given
by Eq.(\ref{mh}) at low momentum and behaves as Eq.(\ref{run})
for a momentum scale larger than the current mass, {\it i.e.} the contribution
of dynamically generated mass to the total mass is negligible for heavy
quarks. For light
quarks ($m_f < \fc^{1/3} $) the self energy is a sum of the dynamical mass
which dominates in the low momentum region but the asymptotic
behavior of self-energy is still given by Eq.(\ref{run}). Only at the chiral limit
$\S$ is fully described by the dynamical mass function. This 
behavior does not change when we consider gluons
with a dynamically generated mass~\cite{paulo}, 
although we obtain smaller values for the dynamical quark 
masses (at $P^2 \rightarrow 0$) in this
case~\cite{papa,napa}. For large gluon masses the signal of chiral
symmetry breaking may disappear when we solve the Schwinger-Dyson
equation in the lowest order~\cite{napa}, but we consider this as
a failure of the method for solving gap equations~\cite{papa} at the
lowest order and still assume that Eqs.(\ref{sope}) and (\ref{mdyn0}) are
valid.

Our ansatz based on the solutions of Ref.~\cite{maro} has the form
\be
\S (P^2) = m_f (P^2) + m_{dyn} (P^2) ,
\l{sfull}
\ee
which is a sum of the bare and dynamical mass. According to
the numerical solutions of Ref.~\cite{maro} and in agreement
with the ansatz already used in Ref.~\cite{castorina,bar} we can
have the following expressions: a) In 
the chiral limit ($cl$)
\be
\S_{cl} (P^2) = m_{dyn} (P^2) ,
\l{clim}
\ee
where
\br
m_{dyn} (P^2) &=& \n^3
\left[ \frac{1}{\n^2} \theta (\n^2 - P^2) \right. \nonumber \\ 
&&+ \frac{1}{P^2} 
\left. \left( \frac{\ln{P^2/\L_Q^2}}{\ln{\n^2/\L_Q^2}} \right)^{\g_m - 1} 
\theta (P^2 -\n^2) \right],
\l{mdf}
\er
where $\n$ is a mass scale which in principle can be related to the
quark condenate, since in QCD, according to Eq.(\ref{sope}), at some 
large mass scale $(\m >> \L_Q)$ we must have, 
$\n^3 \propto (3(N^2 -1)g^2(\m^2)/8N^2) \fc_{\m}$. The above ansatz is equal
to the one used in Ref.~\cite{bar}, where the full effective potential
was computed and $\n$ was considered a variational parameter to be
phenomenologicaly adjusted. In our case we calculate the minimum of energy.
Therefore, our ansatz must approach as close as possible the actual solution
of the Schwinger-Dyson equation. Consequently, to be also compatible with
the low energy phenomenology (Eq.(\ref{mdyn0})), which sets a scale
for all the chiral observables, we set 
\be 
\n \equiv {\fc_{\n}}^{1/3}.
\l{defm}
\ee
b) In the case of a light fermion ($lf$) ( i.e. $m_f < \n$) we
define
\be
m_{lf} (P^2) = m_f (\n^2) \left[ \theta (\n^2 - P^2) +
\left( \frac{\ln P^2/\L_Q^2}{\ln \n^2/\L_Q^2} \right)^{-\g_m} \theta (P^2 -
\n^2) \right], 
\l{mlf}
\ee
which has the correct low and high energy behavior, and use the ansatz
\be
\S_{lf} = m_{lf} (P^2) + m_{dyn} (P^2).
\l{slf}
\ee
The numerical solutions of the Schwinger-Dyson
equations~\cite{maro} show that when the bare mass is much smaller than
the dynamical mass, $\S (P^2)$ starts falling as $1/P^2$ just after 
$P^2 \approx \n^2$, changing to the logarithmic behavior when $\n^3/P^2$ 
becomes of the order of the bare mass. Eq.(\ref{slf}) reproduces this
behavior and can reasonably represent
the masses of the quarks {\it u}, {\it d}, and {\it s} with a suitable choice
of bare masses.
c) For heavy fermions ($hf$) we introduce
\be
m_{hf} (P^2) = m_f  \left[ \theta (m_f^2 - P^2) +
\left( \frac{\ln P^2/\L_Q^2}{\ln m_f^2/\L_Q^2} \right)^{-\g_m} \theta (P^2 -
m_f^2) \right], 
\l{mhf}
\ee
where the basic difference with Eq.(\ref{mlf}) is the choice of arguments 
of the $\theta$ functions. From the results of Ref.~\cite{maro} we could be
tempted to set $\S_{hf} (P^2) = m_{hf}$, which is of course the leading
contribution to $\S (P^2)$ for very heavy quarks.
However, notice that Eq.(\ref{omef}) would give $\< \Omega_f \> = 0$ with this
approximation, and we must stick to Eq.(\ref{sfull}) to obtain the correct
result as will be discussed in the next section.

Eqs.(\ref{sfull})--(\ref{mhf}) give a quite reasonable fit for the full
numerical solutions of the quark Schwinger-Dyson equations. The $\theta$
functions separate the regions where techniques like the OPE and
renormalization group are reliable from the ones where
nonperturbative effects are present and we can at most
rely on phenomenological models.
Being a variational parameter the scale $\n$ can be adjusted in
order to fit observable chiral parameters~\cite{bar}.
Moreover, the freezing of the coupling constant at the scale $\n$
is also implicit in Eq.(\ref{mdf}). Although there
are arguments in the literature showing that this freezing
may already occur at the gluon mass scale~\cite{papa,copa},
we do not expect substantial
differences in our calculation due to this approximation
as will be discussed in the conclusion.

We can now turn to the ansatz for the gluon polarization
tensor. The behavior of $\P$ away from the high momentum
region is much less known and is more controversial. Nonperturbative
solutions for the gluon propagator have been searched for
a long time, but only recently it became more common
to talk about a possible infrared finite behavior of this propagator.
As discussed in Ref.~\cite{mont} there are several
forms proposed in the literature for the gluon polarization
tensor in the infrared, which appear due to different
approximations performed when solving the Schwinger-Dyson
equation for the gluon propagator, and there is
an indication that the vacuum energy is minimized for
gluons with a dynamically generated mass~\cite{mont}. A
strong support for this possibility comes from the gluon 
propagator simulation in lattice QCD~\cite{lat}. These
results may be considered preliminary and large lattice
studies must be pursued, but they definitively
show signals of a gluon mass scale up to reasonably low
momenta~\cite{lat}. We shall assume here that gluons
have indeed a dynamical mass as predicted by Cornwall~\cite{corn}
many years ago, and to obtain an ansatz for the gluon polarization
tensor we are guided by the OPE again. The OPE can teach us
only about the high energy behavior of the dynamical mass which is
given by~\cite{lav}
\be
\P_{OPE} (P^2) \sim - \frac{34 N \pi^2}{9(N^2-1)}
\frac{\gc}{P^2}.  
\l{piuv}
\ee
For the infrared we assume that $\Pi$ freezes at the
value predicted by Eq.(\ref{piuv}) at some scale which
should be determined. This freezing is consistent with
the prediction of Ref.~\cite{corn} of a constant gluon
mass in the infrared. Therefore, we assume
\be
\P (P^2)= \m^2_g \theta (\c \m^2_g-P^2) + \frac{\m^4_g}{P^2}
\theta (P^2-\c \m^2_g)
\l{pian}
\ee
where
\be
\m^2_g = \left( \frac{34 N \pi^2}{9(N^2-1)} \gc \right)^{1/2}.
\l{glmas}
\ee
Notice that $\c$ in Eq.(\ref{pian}) is
a variational parameter to be determined
in our calculation. It defines the
scale that separates the perturbative and nonperturbative 
regions~\cite{pana}. This choice of ansatz is fully based on
the result of the operator product expansion for the gluon
propagator~\cite{lav} and on the phenomenological estimates
of the dynamical gluon mass~\cite{corn,mont}. Note that this
is an ansatz in the case of a pure gluon theory. Of course,
fermions modify this result. 
There are two types of fermionic contributions to
the dynamical gluon mass. One is proportional to the gluon condensate but
appears only at the next order in the coupling constant, i.e. the gluon mass
is going to be modified by a factor $(1 + \alpha_s n_f ...)$.
Therefore, for heavy quarks their
contribution is small due to the smallness of the coupling constant
at the scale of heavy quarks. However, for light fermions this is no
longer true. Unfortunatelly
there is not enough information on the infrared behavior
of the gluon propagator to do better than this. Therefore,
we are forced to consider phenomenologically the contribution of fermions.
The other contribution to
the gluon mass is proportional to the fermion condensate and its mass
$ m_f \<{ \bar{\psi}}\psi\> $, which could be important 
for heavy quarks but as we checked is numerically small compared to the
contribution proportional to the gluon condensate and pure gluon loop.
Eqs.(\ref{sfull}) and (\ref{pian}) are the basic ingredients
of the vacuum energy computation that we present in the next
section.

\section{The gluon self-energy and a relation between the quark and gluon
condensates}        

\noindent

As discussed in the beginning our main goals are to study the
infrared behavior of the gluon propagator and obtain a relation between
the quark and gluon condensates like
Eq.(\ref{qgcon}) through the vacuum energy analysis.
For this we recall that
the vacuum expectation value of the trace of the energy
momentum tensor of QCD is~\cite{cre}
\be
\< \Theta_{\m\m} \> = \frac{\b(g)}{2g} \< G_{\m\nu} G^{\m\nu} \> -
\sum_f  m_f \fc (1 + \g_m),
\l{tmn}
\ee
where the perturbative $\b (g)$ function up to two loops is
\be
\frac{\b(g)}{2g} = - \frac{1}{24} \frac{\a_s}{\pi} 
\left[ (11N-2n_f) + \frac{1}{4} \frac{\a_s}{\pi} 
\left(34N^2-10Nn_f-3n_f \frac{N^2-1}{N}  \right) \right] ,
\l{bfunc}
\ee
with $\a_s = g^2 (\m) / 4\pi$.
Eq.(\ref{tmn}) is related to the vacuum energy through
\be
\< \Omega^{(tr)} \> = \frac{1}{4} \< \Theta_{\m\m} \> .
\l{vt}
\ee
When we compare this last expression for $\< \Omega \>$
with the one of Section 2 we can fix the behavior
of the dynamical gluon mass as well as we can obtain
a relation between the quark and gluon condensates.

The calculation of the vacuum energy using the ansatze of
the previous section is straightforward. However, it is
more convenient to present the results of the calculation 
in different parts. 
The gluon contribution to the vacuum energy is obtained
substituting Eq.(\ref{pian}) into Eq.(\ref{omeg}), and the
result is
\br
\< \Omega_g \> &=& - \frac{3(N^2 -1)}{32\pi^2} 
\Biggl(
\frac{\c -1}{2} - \frac{1}{3} \ln{(\c +1)} - \nonumber \\
&& \left( \frac{1}{6} + \c^2 \right)
\ln{\c} + \left( \frac{3\c^2 + 2}{6} \right) \ln{(\c^2 +1)}
\Biggr)
 \m^4_g ,
\l{romeg}
\er
where $\m_g$ is given by Eq.(\ref{glmas}).

Due to the form of the ansatz for the quark
self-energy its contribution to the vacuum energy has
to be computed only numerically. In the chiral limit we set
$m_0=m_f=0$ into Eq.(\ref{omef}) and easily verify that
\be
\< \Omega_f \>_{cl} = - \frac{N \n^4}{16{\pi}^2}
\left(  2\ln{2} - 1 + 2 \int_1^{\L^2/\n^2} dz \, z \left( \ln (1 + h/z) -
\frac{h/z}{ 1+ h/z}\right)\right) ,
\l{ofchi}
\ee
where $h$ is given by
\be
h = z^{-2} \left( \frac{\ln (z\n^2/\L^2_Q)}{\ln (\n^2/\L^2_Q)}
\right)^{2\g_m-2}. 
\l{defh}
\ee

When we consider bare massive fermions there is an essential modification
in the calculation due to the mass renormalization. We
initially consider the case where $m_f << \n$ and
work with the ansatz of Eq.(\ref{slf}).
Expanding the high energy part in powers of $1/P^2$,
we verify that Eq.(\ref{omef}) reduces to
\br 
\< \Omega_f \>_{lf} &\simeq& - \frac{N}{8{\pi}^2}
\left[ \ \int_0^{\n^2} dz \, z \right.
\left( \ln\frac{z+(\n +m_f)^2}{z+m_f^2} \right.
\left. - \frac{\n (\n +m_f)}{z+\n^2} \right) \nonumber \\ 
&& + m_0 (\L) \n^3 \int_{\n^2}^{\L^2} \frac{dz}{z} 
\left( \ln\frac{z}{\L_Q^2}\right)^{\g_m -1}  \nonumber \\
&& + \int_{\n^2}^{\L^2} dz \, z 
\left. \left( \ln (1 + h \n^2/z) - 
\frac{h\n^2/z}{ 1+ h\n^2/z}\right) \right] \nonumber \\  
&& \simeq -\frac{N}{8{\pi}^2}  \left[ \int_0^{\n^2} dz \, z \right.
\left( \ln\frac{z+(\n +m_f)^2}{z+m_f^2} \right.
\left. - \frac{\n (\n +m_f)}{z+\n^2} \right) \nonumber \\ 
&&+  \frac{\fc_{\n} m_f (\n)}{\g_m} + \int_{\n^2}^{\L^2} dz \, z 
\left. \left( \ln (1 + h\n^2/z) - \frac{h\n^2/z}{ 1+ h\n^2/z}\right)  \right] ,
\l{oflf}
\er
where we kept only the leading
term in $m_f$ in the ultraviolet part as well as the terms independent
of the bare mass. A numerical evaluation of the above expression
shows that $\< \Omega_f \>_{lf} $ has a very small variation with
$m_f$ up to masses of the order of $100 \,\, MeV$ when we
assume $\n \sim 230 \,\, MeV$ with the minimum
becoming $5 \%$ deeper for masses of this order.

For very heavy fermions ($m_f >> \n$) the Schwinger-Dyson equations
tell us that~\cite{maro,paulo} $\S (P^2) \approx m_f (P^2)$, and it
is not difficult to see that the fermion mass effect
is basically erased from the vacuum energy Eq.(\ref{omef}) apart from
a term coming from the mass renormalization, which survives no matter how small
is the contribution from the dynamical mass. Therefore, 
for each heavy fermion of mass $m_f$ we obtain
\be
\< \Omega_f \>_{hf} \simeq -\frac{N \fc_{\n} m_f (\n)}{8\pi^2\g_m}.
\l{ofhf}
\ee
Note that the renormalization group invariant quantity $m_f \fc $ 
appears after the mass renormalization. For light fermions
the terms proportional to the bare mass in $\< \Omega_f \>_{lf} $ are
significantly smaller in comparison with the terms due to the
dynamical mass. For heavy fermions, {\it i.e.} for masses larger than
a few $GeV$, all terms in $\< \Omega_f \>_{hf}$ other than $m_f \fc $ 
can be neglected without affecting considerably our forthcoming numerical
estimates. 

To determine the parameter $\c$ which separates the perturbative and
nonperturbative regions of the gluon self-energy and to
find a relation between the condensates, we consider the
equality 
\be
\< \Omega^{(tr)} \> = \< \Omega_g \> + \< \Omega_f \> .
\l{equ}
\ee
We fix the parameter $\c$ by working with the theory without fermions.
We obtain the following expression:
\br
\frac{\b(g)}{8g} \< G_{\m\nu} G^{\m\nu} \> &=&
- \frac{3(N^2 -1)}{32\pi^2} \m^4_g 
\Biggl(
\frac{\c -1}{2} - \frac{1}{3} \ln{(\c +1)} - \nonumber \\
&& \left( 
\frac{1}{6} + \c^2 \right)
\ln{\c} + \left( \frac{3\c^2 + 2}{6} \right) \ln{(\c^2 +1)}
\Biggr) .
\label{omsg}
\er
Substituting $m_g$ given by Eq.(\ref{glmas}) and considering the
$\b (g)$ function up to one loop we obtain
\be
\left( 
\frac{\c -1}{2} - \frac{1}{3} \ln{(\c +1)} - \left( \frac{1}{6} + \c^2 \right)
\ln{\c} + \left( \frac{3\c^2 + 2}{6} \right) \ln{(\c^2 +1)}
\right) = \frac{11}{34},
\label{echi}
\ee
which gives
\be
\c \approx 0.966797.
\label{valc}
\ee
This value, being so close to $1$, means that the gluon
mass given by Eq.(\ref{glmas}) is a reasonable
scale to separate the perturbative and nonperturbative behaviors
of the gluon propagator. In the following we assume a complete
cancelation of $\< \Omega_g \>$ with the part of $\< \Omega^{(tr)} \>$
coming from the gluonic contribution to the $\b (g)$ function when we
use the above value of $\c$. The phenomenological value of the gluon mass 
obtained in this way is also consistent with previous estimates~\cite{corn}. It
is important to recall that in the OPE calculation of the gluon polarization
tensor fermions do not contribute at leading order to the term
proportional to $\gc$, and this cancelation is indeed complete.

We can now consider the presence of fermions. The case of
heavy quarks is the simplest one. Using Eqs.(\ref{vt}),
(\ref{romeg}), and (\ref{ofhf}) and taking into account the cancelation
mentioned above, for $n_f$ heavy fermions of equal mass we obtain 
\be
\frac{n_f}{48} \gc -
\frac{n_f m_f}{4} \fc (1 + \g_m)
= -\frac{N n_f \fc_{\n} m_f (\n)}{8\pi^2\g_m} ,
\l{ohe}
\ee
entailing
\be
m_f \fc \approx \frac{1}{\kappa} \gc .
\l{fhq}
\ee
Neglecting the second term of the $\b$ function, what is totally
acceptable because $\a_s$ is small on the scale of heavy quarks, we obtain
\be
\kappa = 12 \left( 1+\g_m - \frac{N}{2\pi^2\g_m} \right) ,
\l{k1f}
\ee
which gives 
$\kappa \approx 11.9, \, 12.6, \, 13.2$, respectively for one, two and three
heavy quark flavors. Thus, by calculating the vacuum energy as a function
of the quark and gluon condensates we
have reproduced the old result of Ref.~\cite{shif2}. Note that this
result appears naturally due to the cancelation between the gluonic parts of
the vacuum energy and the fact that we can neglect the heavy quark contribution
to the gluonic part of the effective potential,
because
the heavy quark contribution to the gluon mass proportional to the
gluon condensate appears in the next order of $\alpha_s$.
We would like to note that
the nonleading terms neglected in the operator product
expansion, for example, the gluon condensate contribution to the
quark self-energy $\S \propto m_f \gc /P^4$,  and vice versa the quark 
condensate contribution to the gluon polarization $\P \propto m_f \fc /P^2$
may slightly modify our results, but even if we were intended to
introduce them in the calculation we should remember that we do not have a
reliable phenomenology to correctly describe their behavior at low momenta.

The identity between the two calculations of the vacuum energy can also
provide some information on the gluon propagator in the case of massless
fermions. The physically interesting case is the one where the
bare mass is zero but the chiral symmetry is broken dynamically. It is
known that chiral symmetry breaking happens when the gauge coupling exceeds a
critical value 
\be
\a_c = \frac{\pi}{3 C_2}.
\l{acri}
\ee
In principle this value (or a larger one~\cite{atk,napa}) should be introduced
in the $\b(g)$ in Eq.(\ref{recl}). Unfortunately we do not
know the behavior  of the $\b$ function in this region! In this case we will
use Eq.(\ref{bfunc}) at the extreme of its domain of validity, which is the
best that we can do at the moment. However, we still have another problem
because our ansatz for the gluon polarization is valid only at leading order
in $\a_s$ (we do not explicitly take into account the fermionic
contribution to the gluon polarization tensor,
while we need to compare it to the two-loop $\b$ function).
We proceed in a different way to the case of heavy fermions. For heavy fermions
we used the identity between two ways of calculating the vacuum energy to
obtain a relation between the condensates. Here, vice versa, we use it to
obtain information on the ansatz for the gluon polarization. In what
follows we modify our ansatz for the gluon mass performing the following
replacement in Eq.(\ref{glmas}):
\be
m^4_g \rightarrow m^4_g ( 1 + \a_s n_f c ),
\l{redef}
\ee
where $c$ is another parameter to be determined through Eq.(\ref{equ}). The
parametrization in Eq.(\ref{redef}) was chosen because the 
fermion
contribution to the gluon mass proportional to the gluon condensate
should depend on $n_f$ and $\alpha_s$

In the chiral limit we have
to compute numerically the integral in Eq.(\ref{ofchi}). Assuming that
$\n$ and $\L_Q$ may vary around the values $0.2$ and $0.3$ GeV, we
verified that the integral in this equation contributes at most
$10\%$ of the total value between the brakets of Eq.(\ref{ofchi}),
{\it i.e.} only the low energy region contributes effectively to the vacuum
energy. If we neglect the high energy part of $\< \Omega_f \>$ and
take into account the cancellation already discussed for the gluonic
contributions in Eq.(\ref{equ}) we obtain for the chiral broken phase 
\br
- \frac{1}{96} && \gc 
\left( -2n_f + \frac{\a_s}{4\pi}\left(  34N^2-10Nn_f-3n_f\frac{N^2-1}{N}\right)\right) 
\approx \nonumber \\ 
&&-\frac{3(N^2-1)}{32\pi^2}  \left( \frac{\c -1}{2}  \right.
- \frac{1}{3} \ln{(\c+1)}
- \left( \frac{1}{6} - \c^2 \right)\ln{\c} \nonumber \\ 
&&\left. + \left( \frac{3\c^2 + 2}{6} \right) \ln{(\c^2 +1)} \right) 
\m^4_g (1 + \a_sn_f c ) - \frac{N \n^4 (2\ln{2} -1)}{16{\pi}^2} . 
\l{recl}
\er
With $n_f = 1$, $\a_s \sim \a_c$, $\c$ given by Eq.(\ref{valc}),
$\gc \simeq (0.01) \,\, GeV^4 $~\cite{shif2}, and
$\fc \simeq  (0.23 \, \, GeV)^3 $~\cite{dos}, we obtain
$c \approx 8.7$. This large value of $c$ means that the next order
term in Eq.(\ref{glmas}) in the case of light fermions cannot be neglected. It
would be interesting to compute the next order contribution to the dynamical
gluon mass through the OPE including the effect of fermions and see how it
compares to our estimate. Another numerical relation between the quark and
gluon condensates could also be obtained for the case where the quarks have a
small bare mass, using the values of $\c$ and $c$ determined above and
Eq.(\ref{oflf}) for the fermionic contribution to the effective
potential. However, such relation will be useful only with a
better knowledge of the nonperturbative $\b$ function and of the
next-to-leading order determination of the gluon mass through the
OPE.

\section{Conclusions}         

\noindent

Starting from the effective action for
composite operators~\cite{cornja} we determined the QCD vacuum
energy as a function of the dynamical quark and gluon masses.
The ultraviolet behavior of these masses can be related to the quark and gluon
condensates through the operator product expansion, and the same is 
expected for their infrared parts, although, concerning this low energy
behavior, we do have to rely on phenomenological data. With this procedure we
obtain the vacuum energy described as a function of the condensates and
bare quark masses. Equalizing this expression to the one obtained from the
trace of the energy momentum tensor we were able to fix
the behavior of the dynamical gluon mass. We verified that the gluon
polarization tensor is fairly well represented by the ansatz we have chosen
(Eq.(\ref{pian})) and that the gluon mass given by Eq.(\ref{glmas})
is a reasonable scale to separate perturbative and
nonperturbative behaviors of the gluon propagator. It is remarkable
that our simple ansatz for the dynamical gluon mass gives such a good result.

For heavy fermions we reproduced the old result of QCD sum rules $m_f \fc
\approx \frac{1}{11.9} \gc$~\cite{shif2} which in our case comes naturally from
the effective potential calculation. 
Note that this formula
has been derived at one-loop and is essentially similar to the result of 
Shifman et al. in the sense that it is a consequence of the behavior of
two-point functions. However, the concept of dynamical gluon mass is
relevant here, without it we cannot compute an effective potential free
of ambiguities (as a spurious infrared cutoff).
We also discussed how the ansatz for the
dynamical gluon mass could be improved and how it can be tested with
the next-to-leading order calculation of the OPE.

The method proposed here can be improved in several ways. We certainly
need a description of the $\b$ function in the nonperturbative region,
without this we do not believe that we can go far away 
with these calculations in the light fermion domain. 
The other improvements are related to the
approximations that we have performed to compute $\< \Omega \>$,
{\it i.e.} the choice of the ansatze for quark and gluon propagators.
The computation of  $\< \Omega_g \> $ has shown to be quite
reliable~\cite{mont}, and in spite of neglecting the contribution
of diagrams with quadrilinear vertices and ghosts,
the final result is similar to the Hartree approximation
made by Cornwall in his calculation of the vacuum 
energy~\cite{corn}, and both calculations reduce
the two-loop contributions to a term whose main
contribution to the effective potential is
proportional to $\Pi^2$. In a more rigorous calculation
the two-loop effective potential should be calculated 
including all diagrams, as well as
the full vertices because they are fundamental for
cancelations present in the full calculation~\cite{corn2}.
As we computed the effective potential at the minimum of 
energy we escaped from this very difficult task, and also
minimized the effect of the approximations~\cite{castorina}.

Another point that could be discussed at length is the
question of the dynamical masses ansatze. It is obvious that the
desirable scenario would be a full numerical solution of
the Schwinger-Dyson equations concomitantly with the effective
potential, but this is also an enormous work. The ansatze we
used are in agreement with the phenomenology at low momenta and the
OPE at high momenta. Notice that the important point is
the connection between the dynamical masses and the condensates,
and it is not relevant what mechanism generates the
condensates or masses as long as the relation predicted by the OPE between the
condensates and the masses holds.

Finally, the idea of the calculation is very simple, and we were able
to recover, within the approximations of this method, the relation
between quark and gluon condensates for heavy quarks obtained
by Shifman {\it et al.} many years ago~\cite{shif2}, and verify
that the vacuum energy is well described by a dynamically massive gluon,
whose momentum dependence is displayed in Eq.(\ref{pian}) and
Eq.(\ref{glmas}).

\section*{Acknowledgments}                    

We would like to thank V.P.Gusynin for valuable remarks.
This research was supported by the
Conselho Nacional de Desenvolvimento
Cient\'{\i}fico e Tecnol\'ogico (CNPq) (AAN),
by Fundac\~ao de Amparo \`a Pesquisa do Estado de 
S\~ao Paulo (FAPESP) (EG,AAN) and by Programa de
Apoio a N\'ucleos de Excel\^encia (PRONEX).

\section*{Figure Captions}

\noindent

{\bf Fig.~1} Two-particle irreducible vacuum diagrams contributing
to the effective potential.

{\bf Fig.~2} Diagrams contributing to the gluon polarization tensor.

\newpage

\begin{figure} 
\vskip 1in
\centerline{
\begin{picture}(300,100)(-15,-15)
\put(-35,105){\makebox(0,0)[br]{$V_2(S,D,G)\;\; =$}}
\put(-5,105){\makebox(0,0)[br]{$-\frac{1}{2}$}}
\Gluon(0,110)(75,110){3}{8}
\ArrowArc(37.5,110)(37.5,0,180) 
\ArrowArc(37.5,110)(37.5,180,360) 
\put(145,105){\makebox(0,0)[br]{$-\frac{1}{2}$}}
\Gluon(150,110)(225,110){3}{8}
\DashCArc(187.5,110)(37.5,0,180){3} 
\DashCArc(187.5,110)(37.5,180,360){3} 
\put(-5,0){\makebox(0,0)[br]{$+\frac{1}{6}$}}
\Gluon(0,5)(75,5){3}{8}
\GlueArc(37.5,5)(37.5,0,180){3}{8} 
\GlueArc(37.5,5)(37.5,180,360){3}{8} 
\put(145,0){\makebox(0,0)[br]{$+\frac{1}{8}$}}
\GlueArc(187.5,23.75)(18.75,-90,270){3}{15}
\GlueArc(187.5,-13.75)(18.75,90,450){3}{15}
\end{picture}
}
\label{fig:1}
\vskip 0.5in
\caption{}
\end{figure}
\begin{figure} 
\vskip 1in
\centerline{
\begin{picture}(300,100)(-15,-15)
\put(-35,105){\makebox(0,0)[br]{$\Pi\;\; =$}}
\Gluon(-20,110)(10,110){3}{4}
\Gluon(65,110)(95,110){3}{4}
\ArrowArc(37.5,110)(27.5,0,180) 
\ArrowArc(37.5,110)(27.5,180,360) 
\Gluon(130,110)(160,110){3}{4}
\Gluon(215,110)(245,110){3}{4}
\DashCArc(187.5,110)(27.5,0,180){3} 
\DashCArc(187.5,110)(27.5,180,360){3} 
\put(-25,0){\makebox(0,0)[br]{$-\frac{1}{2}$}}
\Gluon(-20,5)(10,5){3}{4}
\Gluon(65,5)(95,5){3}{4}
\GlueArc(37.5,5)(27.5,0,180){3}{8} 
\GlueArc(37.5,5)(27.5,180,360){3}{8} 
\put(125,0){\makebox(0,0)[br]{$-\frac{1}{2}$}}
\GlueArc(187.5,23.75)(18.75,-90,270){3}{15}
\Gluon(130,5)(187.5,5){3}{7}
\Gluon(187.5,5)(245,5){3}{7}
\end{picture}
}
\label{fig:2}
\vskip 0.5in
\caption{}
\end{figure}
%
%

\end{document}